\documentstyle[w-edbkps,psfig]{w-edbk}

\begin{document}

\tableofcontents

\title[Quantum, classical and semiclassical analyses
of photon statistics in harmonic generation] {Quantum, classical
and semiclassical analyses of photon statistics in harmonic
generation\footnote{To appear in {\em Modern Nonlinear Optics},
ed. M. Evans, {\em Advances in Chemical Physics}, vol. 119(I) 
(Wiley, New York, 2001). This is a part of the chapter on {\em
Nonlinear phenomena in quantum optics} by J. Bajer, M. Du\v{s}ek,
J. Fiur\'asek, Z. Hradil, A. Luk\v{s}, V. Pe\v{r}inov\'a, J.
Reh\'acek, J. Pe\v{r}ina, O. Haderka, M. Hendrych, J. Pe\v{r}ina,
Jr., N. Imoto, M. Koashi, and A. Miranowicz.}}

\markboth{}{Quantum, classical and semiclassical analyses}


\author[J Bajer and A Miranowicz]{JI\v R\'I BAJER$^1$ and ADAM
MIRANOWICZ$^{2,3}$}

\affil{$^{1}$Department of Optics, Palack\'{y} University,
17. listopadu 50, 772~00 Olomouc, Czech Republic}

\vspace{-1.5cm} \affil{$^{2}$CREST Research Team for Interacting
Carrier Electronics, School of Advanced Sciences, The Graduate
University for Advanced Studies (SOKEN), Hayama, Kanagawa
240-0193, Japan}

\vspace{-1.5cm}
\affil{$^{3}$Nonlinear Optics Division, Institute of Physics,
Adam Mickiewicz University, 61-614 Pozna\'n, Poland}

\section{Introduction}
\label{sect1}

Harmonic generation  is one of the earliest discovered and
studied nonlinear optical processes. For 40 years, since the
first experimental demonstration of second-harmonic generation
(SHG) by Franken and co-workers~\cite{Fra61} followed by its
rigorous theoretical description by Bloembergen and
Pershan~\cite{Blo62}, the harmonic generation has unceasingly
been attracting much attention \cite{shgexperiment}. In
particular, harmonic generation has been applied as a source of
nonclassical radiation (see references~\cite{Man95,Per91} for a
detailed account and bibliography). It was demonstrated that
photon antibunched and sub-Poissonian light \cite{Koz77,Kie78}, as
well as second~\cite{Man82} and higher order \cite{Hon85,Hil87}
squeezed light can be produced in SHG. In experimental schemes,
second-harmonic generation is usually applied for the
sub-Poissonian and photon-antibunched light production, whereas
second-subharmonic generation (also referred to as the
\inx{two-photon down conversion}) is used for the squeezed-light
generation \cite{Man95,Per94}. Non-classical effects in
higher-harmonic generation have also been investigated, including
sub-Poissonian \inxx{sub-Poissonian light} photocount
statistics~\cite{Per91,Kie78,Mal91,Baj92},
\inx{squeezing}~\cite{Per91,Koz83,Baj94}, higher-order
squeezing~\cite{Koz86,Du93} according to the Hong-Mandel
definition~\cite{Hon85} or higher-power-amplitude squeezing
\cite{Zha92,Du93} based on Hillery's concept~\cite{Hil87}. In
this contribution, we will study photocount statistics of second
and higher harmonic generations with coherent light inputs.

\inxx{photocount noise} Photocount noise of the observed
statistics can simply be described by the (quantum) {\em Fano
factor} \inxx{Fano factor; quantum}~\cite{Fan47}
\begin{equation}
F^{\rm Q}\equiv\frac{\langle\left(\Delta \hat{n}\right)^2\rangle
}{\left\langle \hat{n}\right\rangle } =\frac{\left\langle
\hat{n}^{2}\right\rangle -\left\langle
\hat{n}\right\rangle^{2}}{\left\langle \hat{n}\right\rangle } ,
\label{N01}
\end{equation}
where $\left\langle \hat{n}\right\rangle$ is the (ensemble) mean
number of detected photons and $\langle\left(\Delta
\hat{n}\right)^2\rangle$ is the variance of photon number. We
also analyze the global (quantum) Fano factor \inxx{Fano
factor; global} defined to be \cite{global}:
\begin{equation}
F^{\rm G}\equiv\frac{\langle\!\langle\left(\Delta
\hat{n}\right)^2\rangle\!\rangle }{\langle\!\langle
\hat{n}\rangle\!\rangle } =\frac{\langle\!\langle
\hat{n}^{2}\rangle\!\rangle -\langle\!\langle
\hat{n}\rangle\!\rangle^{2}}{\langle\!\langle
\hat{n}\rangle\!\rangle } , \label{N02}
\end{equation}
where the mean values $\langle\!\langle\hat{n}^k\rangle\!\rangle$
are obtained by the ensemble and time averaging, i.e.,
\begin{eqnarray}
\langle\!\langle\hat{n}^k\rangle\!\rangle=\lim_{T\rightarrow\infty}
\frac{1}{T}\int_0^T \langle\hat{n}^k(t)\rangle dt. \label{N03}
\end{eqnarray}
In classical trajectory approach, the Fano factor \inxx{Fano
factor; semiclassical} is defined to be
\begin{equation}
F^{\rm S}\equiv \frac{\overline{(\Delta n)^2}
}{\overline{n}}=\frac{\overline {n^{2}}-\overline{n}^{2}}
{\overline{n}}, \label{N04}
\end{equation}
as a semiclassical analogue of the quantum Fano factor. The mean
values $\overline {n^{k}}$ in Eq. (\ref{N04}) are obtained by
averaging over all classical trajectories as will be discussed in
detail in Sects. \ref{sect23} and \ref{sect33}.

Coherent (ideal laser) light has Poissonian photon-number
distribution thus described by the unit Fano factor. For $F<1$,
the light is referred to as \inxx{sub-Poissonian light} {\em
sub-Poissonian} since its photocount noise is smaller than that
of coherent light with the same intensity. Whereas for $F>1$, the
light is called {\em super-Poissonian} \inxx{super-Poissonian
light} with the photocount noise higher than that for coherent
light.

We shall compare different descriptions of photon-number
statistics in harmonic generation within quantum, classical and
semiclassical approaches. First, we will study the exact quantum
evolution of the harmonic generation process by applying numerical
methods including those of Hamiltonian diagonalization and global
characteristics. As a brief introduction, we will show explicitly
that harmonic generation can indeed serve as a source of
nonclassical light. Then, we will demonstrate that the
quasi-stationary sub-Poissonian light can be generated in these
quantum processes under conditions corresponding to the so-called
no-energy-transfer regime known in classical nonlinear optics. By
applying method of classical trajectories, we will demonstrate
that the analytical predictions of the Fano factors are in good
agreement with the quantum results. On comparing second
\cite{Baj99}, third \cite{Baj00} and higher \cite{Baj00a}
harmonic generations in the no-energy-transfer regime, we will
show that the highest noise reduction is achieved in
third-harmonic generation with the Fano-factor of the third
harmonic equal to $F^{\rm Q}_3\approx F^{\rm S}_3=13/16$.

\section{Second-harmonic generation}
\label{sect2} \inxx{harmonic generation; second}

\subsection{Quantum analysis}
\label{sect21}

The quantum process of second-harmonic generation (SHG) can be
described by the following interaction Hamiltonian
\cite{Man95,Per91}:
\begin{equation}
\hat{H}=\hbar g\left( \hat{a}_{1}^{2}\hat{a}_{2}^{\dag
}+\hat{a}_{1}^{\dag 2}\hat{a}_{2}\right) , \label{N05}
\end{equation}
where $\hat{a}_{1}$ and $\hat{a}_{2}$ denote annihilation
operators of the fundamental and second-harmonic modes,
respectively; $g$ is a nonlinear coupling parameter. The
Hamiltonian (\ref{N05}) describes a process of absorption of two
photons at frequency $\omega_1$ and simultaneous creation of a new
photon at the harmonic frequency $\omega_2=2\omega_1$, together
with the inverse process. Unfortunately, no exact solution of
quantum dynamics of the model, described by (\ref{N05}), can be
found. Thus, various analytical approximations or numerical
methods have to be applied in the analysis of the conversion
efficiency, quantum noise statistics or other characteristics of
the process \cite{Per91}. Due to mathematical complexity of the
problem, the investigations of nonclassical effects in harmonic
generation have usually been restricted to the regime of short
interactions (short optical paths or short times). Theoretical
predictions of quantum parameters (including the Fano factor or,
equivalently, the Mandel $Q$-parameter) were obtained under the
\inx{short time approximation} only (see, e.g.,
\cite{Man95,Per91,Baj92}). This is a physically sound
approximation in case of weak nonlinear coupling of optical
fields. The Fano factors \inxx{Fano factor; quantum} under the
short-time approximation (i.e., for $gt\ll 1$) for coherent
inputs $\alpha _{1}=r_{1}\exp \left( i\phi _{1}\right)$ and
$\alpha _{2}=r_{2}\exp \left( i\phi _{2}\right) $ are given by
the expansions (for $r_1,r_2 \neq 0$):
\begin{eqnarray}
F^{\rm Q}_{1} &=&1-4\sin\theta\; r_{2} gt \nonumber \\ &&
+\left\{4r_{1}^{-2} r_{2}^2-2 r_{1}^2+8[2+\cos(2\theta)]
r_{2}^2\right\} (gt)^2 +{\cal O}\{(gt)^3\} ,
\nonumber \\
F^{\rm Q}_{2} &=&1-\frac{16}{3}\sin \theta \;r_{1}^{2}r_{2}(gt)^3
\nonumber \\
&& + \frac{4}{3} \left\{ 2r_{2}^{2}+16r_{1}^{2}r_{2}^{2}-[4+3
\cos(2\theta)]r_{1}^{4} \right\}(gt)^{4} +{\cal O}\{(gt)^5\} ,
\label{N06}
\end{eqnarray}
where $\theta =2\phi _{1}-\phi _{2}$ and ${\cal O}\{x\}$ denotes
the order of magnitude. Eq. (\ref{N06}) determines whether the
generation of harmonics ($\omega +\omega \rightarrow 2\omega$) or
subharmonics ($2\omega \rightarrow \omega +\omega$) occurs. It
also determines the sub-Poissonian or super-Poissonian
photon-number statistics of light generated during the short-time
interactions. For spontaneous SHG process (i.e., for $r_{2}=0$),
the well-known expansions for the quantum Fano factors
\begin{figure}
\vspace*{-3cm} \centerline{\psfig{figure=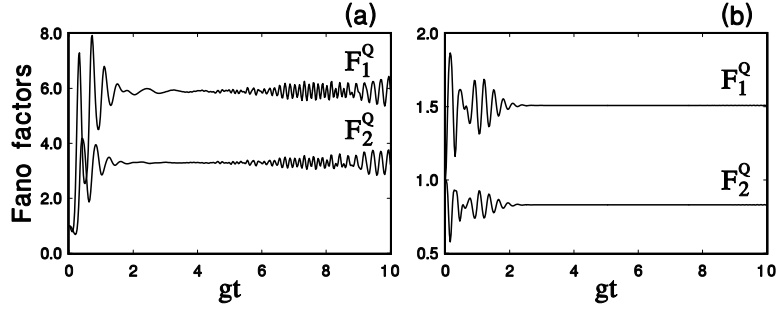,width=15cm}}
\vspace*{-14cm} \caption{Fano factors of the fundamental, $F^{\rm
Q}_{1}$, and the second-harmonic mode, $F^{\rm Q}_{2}$, in the
long-time interaction for initial coherent states with real
amplitudes {\bf (a)} $\alpha _{1}=6, \alpha _{2}=1$, and {\bf
(b)} $\alpha _{1}=6, \alpha _{2}=3$. Case {\bf (a)} is a typical
example of super-Poissonian behavior in both modes outside the
no-energy-transfer regime. In case {\bf (b)}, the harmonic mode
exhibits stable sub-Poissonian statistics with $F^{\rm
Q}_{2}\simeq 0.83$. It is a characteristic example of the
sub-Poissonian behavior within the no-energy-transfer regime along
the line $|\alpha _{1}|=2|\alpha_{2}|$. \label{bajefg01}
}\end{figure}
\vspace*{-8mm}%
\noindent are
\begin{eqnarray}
F^{\rm Q}_{1} &=&1- 2(r_{1}gt)^2 + \frac{4}{3}r_{1}^{2}\left(
3r_{1}^{2}+1\right) (gt)^4+ {\cal O}\{(gt)^6\} ,
\nonumber \\
F^{\rm Q}_{2} &=&1 -\frac{4}{3} (r_{1}gt)^{4}+
\frac{4}{45}r_{1}^{4}\left( 36r_{1}^{2}+17\right) (gt)^{6} +{\cal
O}\{(gt)^8\}, \label{N07}
\end{eqnarray}
or, equivalently, for the normally-ordered photon-number variances
\cite{Koz77,Kie78,Man95}:
\begin{eqnarray}
\langle:\left(\Delta \hat{n}_1\right)^2:\rangle &\equiv&
\langle\left(\Delta \hat{n}_1\right)^2\rangle -\left\langle
\hat{n}_1\right\rangle = - 2r_{1}^4 (gt)^2 +{\cal O}\{(gt)^4\} ,
\nonumber \\
\langle:\left(\Delta \hat{n}_2\right)^2:\rangle&\equiv&
\langle\left(\Delta \hat{n}_2\right)^2\rangle -\left\langle
\hat{n}_2\right\rangle = -\frac{4}{3} r_{1}^8(gt)^{6}+{\cal
O}\{(gt)^8\}. \label{N08}
\end{eqnarray}
It is seen that the photon-number statistics of fundamental mode
exhibits, in the short-time regime, much stronger sub-Poissonian
behavior than that of harmonic mode.

For longer interaction times ($gt>1$), there are no exact
analytical solutions, thus the numerical analysis has to be
applied. We have used two methods to study the quantum dynamics:
(i) the well-known Hamiltonian diagonalization proposed by Walls
and Barakat \cite{Wal70} and (ii) the method of global
characteristics based on manipulation with spectra \cite{global}.
These methods can be applied for arbitrary initial photon
statistics. Nevertheless, for the purpose of our paper, we
restrict our analysis to the initial coherent fields solely. Due
to computational difficulties, the results can be obtained for
small numbers of interacting photons only. The analysis of about
one hundred interacting photons reaches practically the
computational capabilities of the standard mathematical software.

\begin{figure}
\vspace*{0cm}
\centering
\centerline{\psfig{figure=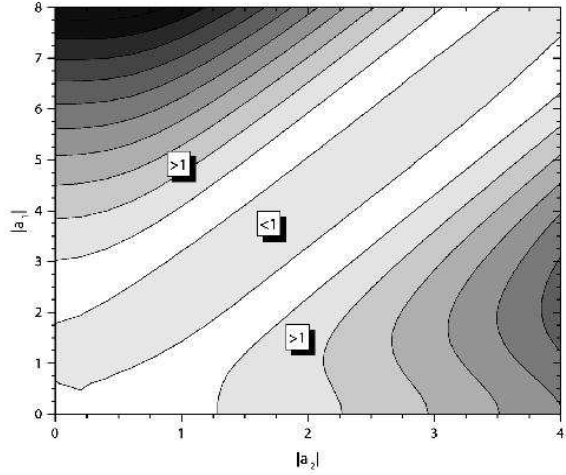,width=7.5cm}}
\vspace*{0cm}
\caption{\inxx{Fano factor; global}
Global Fano factor, $F^{\rm G}_{2}$, of the
second-harmonic mode as a function of the initial coherent state
amplitudes $\alpha _{1}$ and $\alpha _{2}$ with $\theta =0$. It
is seen that the harmonic mode exhibits globally sub-Poissonian
behavior ($F^{\rm G}_{2}<1$) near the diagonal $|\alpha _{1}|=2|\alpha
_{2}|$ and $\theta=0$. The darker region the higher value of $F^{\rm
G}_{2}$. The counter lines are drawn at $F^{\rm
G}_{2}=1,1.5,\cdots,5.5.$
\vspace*{-0.8cm} } \label{bajefg02} \end{figure}

\begin{figure}[h]
\vspace*{-0.6cm}
\centerline{\psfig{figure=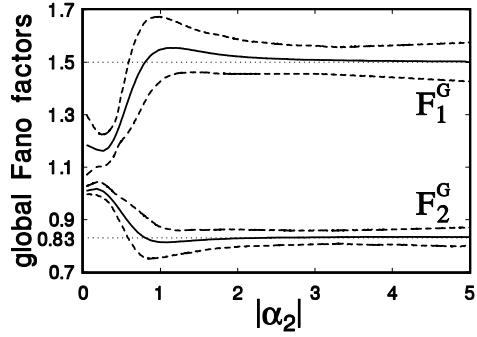,width=2.7in}} \vspace*{0cm}
\caption{\inxx{Fano factor; global} Global Fano factors $F^{\rm
G}_{1}$ and $F^{\rm G}_{2}$ along the line $|\alpha
_{1}|=2|\alpha _{2}|$ for $\theta =0$. Dotted lines denote the
RMS deviation of oscillations in the long-time interaction. It's
seen that the harmonic mode exhibits globally sub-Poissonian
behavior. \label{bajefg03} \vspace*{-0.6cm} }\end{figure}

Analysis of a typical evolution of the Fano factors $F^{\rm
Q}_{1,2}$, such as presented in Fig. \ref{bajefg01}(a), leads to
the conclusion that after initial short-time ($gt<1$) relaxations
in both modes, a strongly super-Poissonian ($F^{\rm Q}_{1,2}\gg
1$). This behavior occurs for the majority of initial coherent
states $|\alpha _{1}\rangle$ and $|\alpha _{2}\rangle$ except a
certain set of initial states concentrated along the line
$|\alpha _{1}|=2|\alpha _{2}|>0$ and $\theta \simeq 0$ [see Fig.
\ref{bajefg01}(b)]. The same conclusion can be drawn by analyzing
the global Fano factors $F^{\rm G}_{1,2}$. We find that the
global Fano factor of the harmonic mode remains independent of
amplitude $|\alpha _{k}|$ and equal to $F^{\rm G}_{2}=0.83<1$
along the line $|\alpha _{1}|=2|\alpha _{2}|$ (see Figs.
\ref{bajefg02} and \ref{bajefg03}). As depicted in Fig.
\ref{bajefg01}(b), when the initial relaxation oscillations fade
out, the harmonic mode remains sub-Poissonian for a long
interaction time interval. In the classical theory of SHG, this
case is referred to as the \inx{no-energy-transfer regime}
\cite{Jex97}, because of the conservation of energy in every
mode. We have found a quantum analog of this regime for coherent
inputs with amplitudes satisfying the conditions: $|\alpha
_{1}|=2|\alpha _{2}|$ and $\theta \simeq 0$.

\begin{figure}[h]
\hspace*{-3mm}
\centerline{\psfig{figure=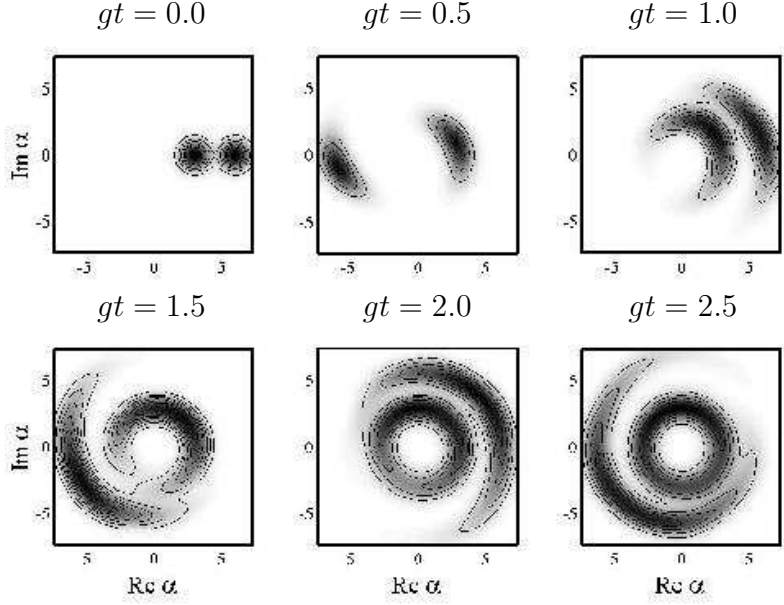,width=10.5cm}}
\vspace*{-.4cm} \caption{Quantum evolution of the $Q$-function for
the fundamental (outer contour plots) and the second-harmonic mode
(inner plots) at six time moments for initial coherent states
with $\alpha_{1}=6,\alpha _{2}=3, \theta =0$. Solution obtained by
quantum numerical method.} \vspace*{-.2cm} \label{bajefg04}
\end{figure}

By analyzing Figs. \ref{bajefg02} and \ref{bajefg03}, an
intriguing question arises: Why does the harmonic-field photocount
statistics in the no-energy-transfer regime remain sub-Poissonian
with the Fano factor almost independent of the interaction time
$gt>1$? This behavior can be understood better by plotting the
Husimi $Q$-function. Let us have a look at the snapshots of
typical evolution of $Q$-functions for both modes at six times
$gt=0,0.5\cdots,2.5$ with initial amplitudes $\alpha
_{1}=6,\alpha _{2}=3$ (Fig. \ref{bajefg04}). These results were
obtained numerically and represent the exact quantum solution of
the model (\ref{N05}). One can observe how the cross-sections of
the $Q$-functions change from circles (for initially coherent
fields) through crescents into rings. We note that both modes have
relatively small photon-number variances and small Fano factors,
$F^{\rm Q}_{1}\approx 1.50$ and $F^{\rm Q}_{2}\approx 0.83$ [see
also Fig. \ref{bajefg01}(b)]. The ring shapes, once formed, are
very stable. So, not only the Fano factors, but the entire quantum
states become stationary.

The $Q$-functions are very wide, thus no linearization of the
quantum problem is possible and no pure quantum technique can be
used for estimation of the observed values $F^{\rm Q}_{1}\approx
1.50$ and $F^{\rm Q}_{2}\approx 0.83$. However, good quantitative
explanation of these numerical values can be obtained by the
method of  classical trajectories as will be shown in Sect.
\ref{sect23}.

Our discussion is focused on photon-number statistics rather than
squeezing or other phase-related properties. Nevertheless, by
analyzing the $Q$-function evolution presented in Fig.
\ref{bajefg04}, we can draw the conclusion that \inx{squeezing}
cannot be observed for initial coherent fields at interaction
times exceeding the relaxation time. In fact, the quadrature
squeezing variances ($k=1,2$)
\begin{eqnarray}
S^{\rm Q}_k \equiv\langle(\Delta \hat{X}_k)^2\rangle =
\langle[\Delta(\hat{a}_k {\rm e}^{-{\rm
i}\theta}+\hat{a}_k^{\dagger} {\rm e}^{{\rm i}\theta})]^2\rangle
\label{N09}
\end{eqnarray}
are monotonically rising from the standard shot-noise-limit
($S^{\rm Q}_k=1$) to much more noisy state with the saturated
quadrature variances
\begin{eqnarray}
S^{\rm Q}_1 = 1+8r^{2}\gg 1,
\nonumber\\
S^{\rm Q}_2 = 1+2r^{2}\gg 1. \label{N10}
\end{eqnarray}
It is evident that the no-energy-transfer regime is not useful
for the quadrature squeezing generation.

\subsection{Classical analysis}
\label{sect22}

Complete quantum solution of the model given by Hamiltonian
(\ref{N05}) can be found by applying sophisticated numeric methods
on a fast computer only. However, since we \linebreak\pagebreak
\vspace*{.2cm}

\vspace*{-9.8cm}
\hspace*{1.3cm}{\large $gt=0.0$} %
\hspace*{2.0cm}{\large $gt=0.5$} %
\hspace*{2.0cm}{\large $gt=1.0$} %

\vspace*{3.4cm}
\hspace*{1.3cm}{\large $gt=1.5$} %
\hspace*{2.0cm}{\large $gt=2.0$} %
\hspace*{2.0cm}{\large $gt=2.5$} %
\vspace*{5.5cm}

\noindent are interested in a special type of solution for strong
fields, we can adopt approximate classical and semiclassical
methods to obtain some analytical results.

In analogy with Eq. (\ref{N05}), the classical model of SHG can
be described by
\begin{equation}
{\cal H}= g\left[
\alpha_{1}^{2}\alpha_{2}^{*}+(\alpha_{1}^*)^{2}\alpha_{2}\right] ,
\label{N11}
\end{equation}
where $\alpha_{1}$ and $\alpha_{2}$ are complex amplitudes of the
fundamental and second-harmonic modes, respectively, and $g$ is a
nonlinear coupling parameter. The exact solution of the model,
described by (\ref{N11}), is well-known (see, e.g., \cite{Boy91}).
The solution is periodic and can be written in terms of the
Jacobi elliptic function. A few special cases (e.g., the
phase-matched second-harmonic generation) have monotonous
solution described by hyperbolic functions. The classical
solution is a good approximation for strong fields, for which
gives correct predictions of the output light intensities and
frequency-conversion efficiency. Unfortunately, it cannot be used
to describe the photocount noise and other statistical properties
of generated light. Now, we will summarize some classical
results, which we will be used in the method of classical
trajectories.

The Hamiltonian (\ref{N11}) for the classical SHG leads to the
following system of complex differential equations \cite{Boy91}
\begin{eqnarray}
\dot{\alpha}_{1}&=&-2{\rm i}g\alpha _{1}^{\ast }\alpha _{2},
\nonumber\\
\quad \dot{\alpha} _{2}&=&-{\rm i}g\alpha _{1}^{2}. \label{N12}
\end{eqnarray}
One obtains, after substitution of $\alpha _{k}=r_{k}{\rm e}^{{\rm
i}\phi _{k}}$, a new system of real equations for the amplitudes
and phases:
\begin{eqnarray}
\dot{r}_{1}&=&-2r_{1}r_{2}\sin \theta ,
\nonumber\\
\dot{r}_{2}&=&r_{1}^{2}\sin \theta ,
\nonumber\\
\dot{\theta}&=&\left( r_{1}^{2}/r_{2}-4r_{2}\right) \cos \theta ,
\label{N13}
\end{eqnarray}
where $\theta =2\phi _{1}-\phi _{2}.$ The system has two
integrals of motion: $E=r_{1}^{2}+2r_{2}^{2}=n_{1}+2n_{2}$ and
$\Gamma =r_{1}^{2}r_{2}\cos \theta$. By extracting $r_{1}$ and
$\theta $ from Eq. (\ref{N13}), we get the following equation for
$r_{2}$:
\begin{equation}
\left( r_{2}\dot{r}_{2}/g\right) ^{2}+\Gamma ^{2}=r_{2}^{2}\left(
E-2r_{2}^{2}\right) ^{2}, \label{N14}
\end{equation}
or even in simpler form for the intensity $n_{2}=r_{2}^{2}$:
\begin{equation}
\left( \dot{n}_{2}/2g\right) ^{2}+\Gamma ^{2}=n_{2}\left( E-2n_{2}\right)
^{2}.
\label{N15}
\end{equation}
Separation of $t$ and $n_{2}$ leads to the equation
\begin{equation}
2g{\rm d}t=\frac{{\rm d}n_{2}}{\sqrt{n_{2}\left( E-2n_{2}\right)
^{2}-\Gamma ^{2}}},
\label{N16}
\end{equation}
which can be rewritten as
\begin{equation}
4g{\rm d}t=\frac{{\rm d}n_{2}}{\sqrt{\left( a-n_{2}\right) \left(
b-n_{2}\right) \left( n_{2}-c\right) }}.
\label{N17}
\end{equation}
where the numbers $a,b,c$ are the roots of cubic equation
$n_{2}\left( E-2n_{2}\right) ^{2}-\Gamma ^{2}=0$. For $c\leq
u\leq b<a$, the solution of Eq. (\ref{N17}) reads as
\begin{equation}
\int_{c}^{u}\frac{dx}{\sqrt{\left( a-x\right) \left( b-x\right)
\left( x-c\right) }}=\allowbreak \frac{2}{\sqrt{a-c}}\, {\rm
asn}\!\left( \sqrt{ \frac{u-c}{b-c}},k\right) \label{N18}
\end{equation}
in terms of the inverse Jacobi elliptic function, ${\rm asn}\left(
x,k\right)$, with parameter $k=\sqrt{\frac{b-c}{a-c}}$. Finally,
the inversion of (\ref{N18}), gives the required solution
\begin{eqnarray}
n_{2}\left( t\right) =c+\left( b-c\right) \, {\rm
{sn}}^{2}\!\left[ 2g\sqrt{a-c} \left( t-t_{0}\right) ,k\right]
{\rm ,} \label{N19}
\end{eqnarray}
where ${\rm {sn}}(u,k)$ is the \inx{Jacobi elliptic function}
with the same parameter $k$. Solution (\ref{N19}) can be
simplified in special cases. In particular, the well-known
elementary solution is obtained for second-harmonic generation
from vacuum, where $r_{1}\left( 0\right) =r$ and $r_{2}\left(
0\right) =0.$ In this case $k=1$ and the Jacobi elliptic function
simplifies to hyperbolic tangent. Thus, the solution reads as
\begin{eqnarray}
r_{1}\left( t\right) &=&r\cosh (\sqrt{2} rgt),
\nonumber\\
r_{2}\left( t\right) &=& \frac{r}{\sqrt{2}}\tanh( \sqrt{2}rgt)
\label{N20}
\end{eqnarray}
and $\theta \left( t\right) =\pi /2.$ Subharmonic generation does
not occur in this classical model, since for $r_{1}\left( 0\right)
=0$ and $r_{2}\left( 0\right) =r$ implies that $r_{1}\left(
t\right) =0,\;r_{2}\left( t\right) =r$ for any evolution time $t$.
Another important special case of solution (\ref{N19}) can be
obtained for the initial zero phase difference, $\theta \left(
0\right) =0$, and the initial amplitudes satisfying $r_{1}\left(
0\right) =2r$ and $r_{2}\left( 0\right) =r$. Here,
$E=6r^{2},\;\Gamma =4r^{3},\;a=4r^{2},\;b=c=r^{2},\;k=0$ and
Jacobi elliptic function simplifies to trigonometric sinus.
Finally, this elementary solution reads as
\begin{eqnarray}
\alpha _{1}\left( t\right) &=&2re^{-2{\rm i}rgt},
\nonumber\\
\alpha _{2}\left( t\right) &=&re^{-4{\rm i}rgt}, \label{N21}
\end{eqnarray}
which corresponds to the \inx{no-energy-transfer regime}, in
which energy is conserved in every mode. Phase trajectories of
that solution are presented in Fig. \ref{bajefg05}(a). The
slightly perturbed solution in the no-energy-transfer regime can
also be approximated by $k\approx 0$ and elementary function
sinus with small amplitude $\left( b-c\right) \approx 0$ [see Fig.
\ref{bajefg05}(b),(c)].

\subsection{Classical trajectory analysis}
\label{sect23}
\inxx{classical trajectory analysis}

The answer to our question concerning the origin of sub-Poissonian
behavior can be found by the method of classical trajectories.
The method is very general. It can be applied in the analysis of
almost every nonlinear quantum process. Even external pumping and
energy losses can be easily described. In the classical trajectory
approach to SHG \cite{Nik91}, deterministic solutions of the
classical SHG are used, while quantum noise  of initial fields is
artificially simulated by Gaussian distribution. One can study
the time evolution of the bunch of trajectories like the
evolution of quantum distributions. This semiclassical method can
often shed some light on complicated quantum dynamics. For strong
inputs, where the quantum noise can be assumed small, the method
gives surprisingly good results.

According to the classical trajectory method, one assumes that the
input stochastic amplitudes are of the form
\begin{eqnarray}
\alpha _{1}&=&r_{1}+x_{1}+{\rm i}y_{1},
\nonumber\\
\alpha _{2}&=&r_{2}+x_{2}+{\rm i}y_{2}, \label{N22}
\end{eqnarray}
where $r_{k}$ are coherent complex amplitudes, whereas $x_{k}$ and
$y_{k}$ are real and mutually independent Gaussian stochastic
quantities with identical variances
\begin{equation}
\sigma ^{2}=1/4\ll r_{k}^{2} \label{N23}.
\end{equation}
\begin{figure}
\vspace*{-5mm} \centerline{\psfig{figure=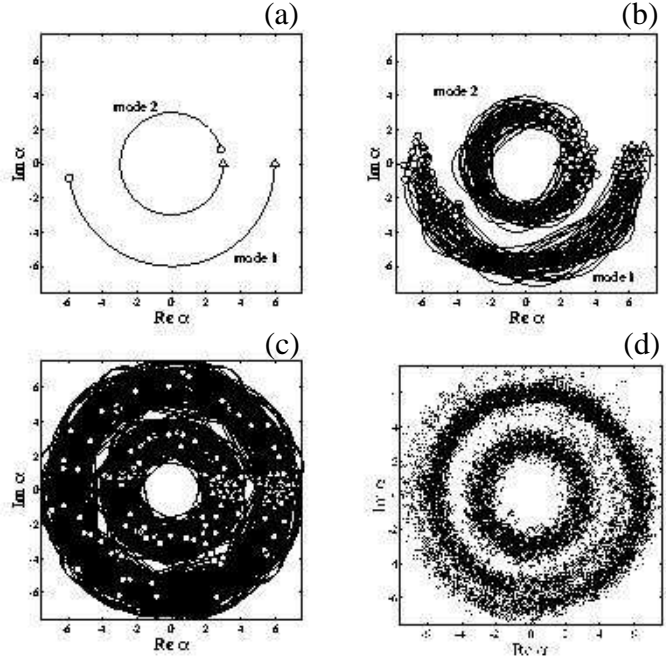,width=9cm}}%
\vspace*{-5mm} \caption{ Classical trajectories of the fundamental
and second-harmonic modes in the no-energy-transfer regime for
$\alpha _{1}=6$ and $\alpha _{2}=3$: {\bf (a)} classical
evolution according to Eq.  (\protect\ref{N19}) for $0<gt<0.5$;
{\bf (b)} fifty random trajectories out of 10,000 trajectories
used in the simulations for $0<gt<0.5$; {\bf (c)} same as in (b),
but for $0<gt<5$; {\bf (d)} snapshot of the $Q$-function,
obtained from 10,000 random trajectories at time $gt=5.0$.
Triangles denote starting points ($gt=0$) of the trajectories and
circles are their ends. \vspace*{-0.5cm}
\label{bajefg05} }\end{figure}%

\vspace*{-12.3cm}
\hspace*{4.4cm} {\large (a)} %
\hspace*{4.2cm}{\large (b)}

\vspace*{3.9cm}
\hspace*{4.5cm}{\large (c)}%
\hspace*{4.3cm}{\large (d)}%
\vspace*{7.6cm}

\begin{figure}
\vspace*{-4.5cm}
\centerline{\psfig{figure=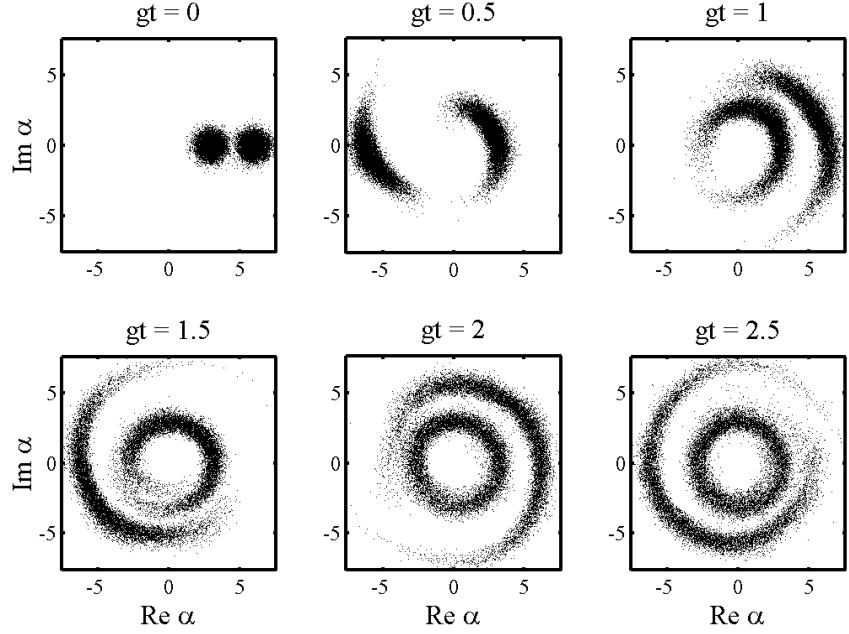,width=14cm}}
\vspace*{-5cm}
\caption{Classical trajectory simulation of quantum evolution of
the $Q$-function for the same initial conditions and interaction
times as in Fig. \protect\ref{bajefg04}. In our simulation 10,000
trajectories were calculated.
}
\label{bajefg06}
\end{figure}

\noindent By analogy with our quantum analysis, we calculate the
semiclassical Fano factor, defined by Eq. (\ref{N04}), and
quadrature \inx{squeezing} variance
\begin{eqnarray}
S^{\rm S}_k \equiv \overline{(\Delta X_k)^2} =
\overline{[\Delta(\alpha_k {\rm e}^{-{\rm
i}\theta}+\alpha_k^*{\rm e}^{{\rm i}\theta})]^2} \label{N24}
\end{eqnarray}
as counterparts of quantum parameters (\ref{N01}) and
(\ref{N09}), respectively. By applying the method of classical
trajectories with the noise variance given by Eq. (\ref{N23}), we
find the semiclassical quadrature squeezing and Fano factor
\inxx{Fano factor; semiclassical} given by:
\begin{eqnarray}
S^{\rm S}_k &=&4\sigma ^{2}=1,
\nonumber\\
F^{\rm S}_{k}&\approx& 4\sigma ^{2}=1, \label{N25}
\end{eqnarray}
respectively. According to the described method, one needs to
solve thousands of the classical SHG trajectories. The mean
values are simply obtained by averaging over all these
trajectories. In Fig. \ref{bajefg06}, we have presented
graphically snapshots in a selected time-interval of all complex
solutions in phase space. These clouds of points naturally
correspond to the $Q$-functions in the quantum picture (see Fig.
\ref{bajefg04}). We have found that this semiclassical method
gives the results surprisingly similar to the quantum results
even for relatively weak fields! This very good agreement is
clearly seen by comparing Figs. \ref{bajefg04} and \ref{bajefg06},
where the initial amplitudes are chosen to be $\alpha _{1}=6,$
$\alpha _{2}=3$. The patterns given by fifty random trajectories
out of the total number of 10,000 analyzed trajectories are shown
in Fig. \ref{bajefg05}(b) in the time interval $gt\in \left(
0,0.5\right) $ and Fig. \ref{bajefg05}(c) in $gt\in \left(
0,5\right)$. The final snapshot of the ``cloud'' ring at $gt=5$
is given in Fig. \ref{bajefg05}(d).

The method of classical trajectories can be used not only
numerically (Figs. \ref{bajefg05} and \ref{bajefg06}) but also
analytically in special cases.  For example, the evolution of
low-noise fields in the no-energy-transfer regime can be found
analytically in the first approximation with the solution given
by elementary trigonometric functions. To show this, let us
analyze integrals of motions. On assuming the initial amplitudes
of the form $\alpha _{1}=2r+x_{1}+iy_{1}$ and $\alpha
_{2}=r+x_{2}+iy_{2}$, the integrals of motion can be expressed in
the form of successive corrections
\begin{eqnarray}
E &=&6r^{2}+\Delta E_{1}+\Delta E_{0,}
\nonumber \\
\Gamma &=&4r^{3}+\Delta \Gamma _{2}+\Delta \Gamma _{1}+\Delta \Gamma _{0},
\label{N26}
\end{eqnarray}
where we denote $\Delta E_{1}=4r\left( x_{1}+x_{2}\right)$,
$\Delta E_{0}=x_{1}^{2}+y_{1}^{2}+2\left(
x_{2}^{2}+y_{2}^{2}\right) $, and $\Delta \Gamma _{2}=4r^{2}\left(
x_{1}+x_{2}\right) ,$ $\Delta \Gamma _{1}=r\left(
x_{1}^{2}-y_{1}^{2}+4x_{1}x_{2}\right) $ and $\Delta \Gamma
_{0}=x_{2}\left( x_{1}^{2}-y_{1}^{2}\right) +2x_{1}y_{1}y_{2}.$
By substituting $n_{2}=E/6+\epsilon$, where $\epsilon $ is a
small correction, and after omitting the cubic term $2\epsilon
^{3}$, we find that the denominator in Eq. (\ref{N16}) can be
approximated by the quadratic function
\begin{equation}
n_{2}\left( E-2n_{2}\right) ^{2}-\Gamma ^{2}\allowbreak \approx \allowbreak
\allowbreak 2E\left( A^{2}-\epsilon ^{2}\right) .
\label{N27}
\end{equation}
Now, we can perform integration of these elementary functions
leading to the simple result
\begin{equation}
n_{2}\left( t\right) =r^{2}+B+A\cos \Omega gt,
\label{N28}
\end{equation}
where $\Omega =\sqrt{8E},$ $A=\frac{2}{3}r\sqrt{\left(x_{1}-2x_{2}
\right) ^{2}+3\left( y_{1}^{2}+y_{2}^{2}\right) }$ and
$B=\frac{2}{3} r\left( x_{1}+x_{2}\right) $.  We get a similar
result
\begin{equation}
n_{1}\left( t\right) =E-2n_{2}=4r^{2}+4B-2A\cos \Omega gt
\label{N29}
\end{equation}
for the fundamental (or subharmonic) mode. Both solutions are
constant functions weakly perturbed by harmonic function. The
evolution in phase space can be understood clearly by analyzing
Figs. \ref{bajefg05} and \ref{bajefg06}. Due to the frequency
dispersion $\Omega \left( \{x_{k},y_{k}\}\right) $ [see Fig.
\ref{bajefg05}(b),(c)], different trajectories are drifting
variously and create a crescent-shape cloud in phase space, which
develops later into a full ring as seen in Figs. \ref{bajefg05}
and \ref{bajefg06}. One has to perform the averaging of solutions
to calculate the required statistical moments. We find
$\overline{n}_{1}=4r^{2},$ $\overline{n}_{2}=r^{2},$ and
\begin{eqnarray}
\overline{n_{1}^{2}}&=&16r^{4}+16\overline{B^{2}}+2\overline{A^{2}},
\nonumber\\
\overline{n_{2}^{2}}&=&r^{4}+\overline{B^{2}}+\frac{1}{2}\overline{A^{2}},
\label{N30}
\end{eqnarray}
where $\overline{B}=0,$ $\overline{A^{2}}=\allowbreak
\frac{44}{9}r^{2}\sigma ^{2}=\frac{11}{9} r^{2}$,
$\overline{B^{2}}=\frac{8}{9}r^{2}\sigma ^{2}=\frac{2}{9}r^{2}$
and $\overline{\cos ^{2}\Omega gt}=\frac{1}{2}$. Finally, we
arrive at the semiclassical Fano factors \inxx{Fano
factor; semiclassical} given by simple rational numbers:
\begin{eqnarray}
F^{\rm S}_{1} &=& \frac{1}{r^{2}}\left(
4\overline{B^{2}}+\frac{1}{2}\overline{A^{2}}\right) =\frac{3}{2},
\nonumber \\
F^{\rm S}_{2} &=& \frac{1}{r^{2}}\left(
\overline{B^{2}}+\frac{1}{2}\overline{A^{2}}\right) =\frac{5}{6}.
\label{N31}
\end{eqnarray}
By analyzing Figs. \ref{bajefg01}(b) and \ref{bajefg03} as well as
tables \ref{bajetb01} and \ref{bajetb02}, we conclude that our
estimations (\ref{N31}) are in very good agreement with those
values of Fano factors obtained by the quantum numerical analysis
of Sect. \ref{sect22}.

\newpage
\section{Higher-harmonic generation}
\label{sect3} \inxx{harmonic generation; higher}

\subsection{Quantum analysis}
\label{sect31}

In this section, we will generalize our results of Sect.
\ref{sect2} to describe the processes of the $N$-th harmonic
generation. Again, we will focus on predictions of the
sub-Poissonian photon-number statistics.

Processes of the $N$-th harmonic or subharmonic generation can be
described by the conventional interaction Hamiltonian (e.g.,
\cite{Per91})
\begin{equation}
\hat{H}=\hbar g\left( \hat{a}_{1}^{N}\hat{a}_{N}^{\dag
}+\hat{a}_{1}^{\dag N}\hat{a}_{N}\right) \label{N32}
\end{equation}
for $N=2,3,\cdots $. In (\ref{N32}), $\hat{a}_{1}$ and
$\hat{a}_{N}$ denote annihilation operators of the fundamental
and $N$-th harmonic modes, respectively, and $g$ is a nonlinear
coupling parameter.  For short evolution times, \inxx{short time
approximation} the following approximation of the quantum Fano
factors \inxx{Fano factor; quantum} can be obtained for the
fundamental mode \cite{Baj92}
\begin{eqnarray}
F^{\rm Q}_{1} &=&1-2N\left( N-1\right) r_{1}^{N-2}r_{2}\sin
\theta\; gt+{\cal O}\{(gt)^2\} \label{N33}
\end{eqnarray}
with $N=2,3,...$, and for higher harmonics:
\begin{eqnarray}
F^{\rm Q}_{3} &=&\allowbreak 1-36r_{1}^{3}r_{2}\left(
r_{1}^{2}+2\right) \sin \theta\;(gt) ^{3}  +{\cal O}\{(gt)^4\},
\nonumber \\
F^{\rm Q}_{4} &=&1-64r_{1}^{4}r_{2}\left(
17+12r_{1}^{2}+2r_{1}^{4}\right) \sin \theta\;(gt) ^{3} +{\cal
O}\{(gt)^4\}, \label{N34}
\end{eqnarray}
where $r_{k}$ are input amplitudes, and $\theta =N\phi _{1}-\phi
_{N}$ is the input phase mismatch. For spontaneous harmonic
generation  (i.e., for $r_{N}=0$), Eqs. (\ref{N33})-(\ref{N34})
simplify to the formulas derived by Kozierowski and Kielich
\cite{Koz83}. This analysis shows the possibility of
sub-Poissonian light generation in short-time regime under the
proper phase condition.

On testing different coherent input amplitudes and phases in
order to minimize the Fano factor for long-interaction times, we
have discovered a regime, for which the harmonic field exhibits
the quasi-stationary sub-Poissonian photocount noise. The regime
occurs if the ratio of amplitudes $|\alpha _{1}|$ and $|\alpha
_{N}|$ is equal to $N$, and phases are related by $N\phi
_{1}=\phi _{N}$. As described in Sect. \ref{sect2} for SHG, this
is a quantum analog of the \inx{no-energy-transfer regime}
\cite{Jex97} known from classical nonlinear optics as an
evolution exhibiting the no-energy transfer between the
interacting modes. The intensities of both modes remain
quasi-stationary during the interaction. Obviously, in quantum
analysis some small energy fluctuations between modes are
observed as a consequence of vacuum fluctuations. However, the
influence of energy fluctuations can be neglected for strong
fields.
\begin{figure}
\vspace*{-3cm}
\centerline{\psfig{figure=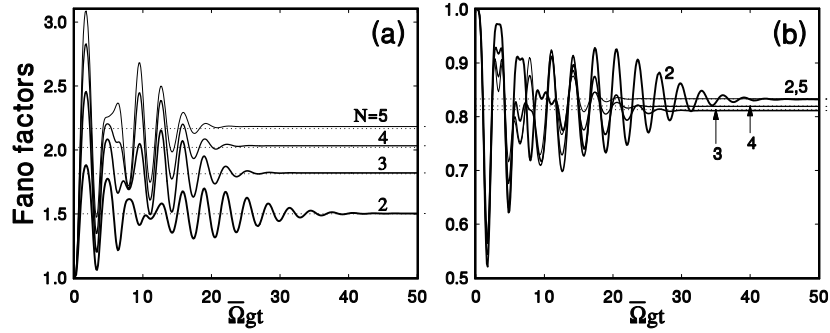,width=16cm}}
\vspace*{-15cm}
\caption{ Time evolution of the exact quantum Fano factors: {\bf
(a)} $F^{\rm Q}_{1}=F^{\rm Q}_{1}(N)$ for the fundamental mode
and {\bf (b)} $F^{\rm Q}_{N}$ for the harmonic mode in
$N$th-harmonic generation for $N$=2 (thickest curve), 3, 4, and 5
(thinnest curve). Time $t$ is rescaled with frequency
$\overline{\Omega }$, given by (\ref{N52}), and coupling constant
g. The harmonic-mode amplitude is $r=r_{N}=5$. The dotted lines
correspond to the semiclassical Fano factors, given by
(\ref{N54}) and (\ref{N55}). It is seen that the fundamental mode
is super-Poissonian, whereas the harmonic mode is sub-Poissonian
for all non-zero evolution times. \label{bajefg07} }\end{figure}

\begin{figure}
\vspace*{-3.3cm}
\centerline{\psfig{figure=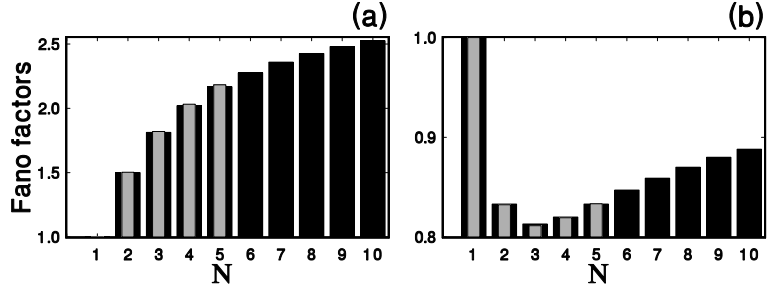,width=15cm}}
\vspace*{-14.5cm} \caption{ Semiclassical (solid bars) and quantum
(dithered bars) Fano factors versus order $N$ of harmonic
generation for {\bf (a)} fundamental and {\bf (b)} $N$th-harmonic
modes in the quasi-stationary no-energy-transfer regime. Figures
(a) and (b) for $N=1-5$ correspond to tables 1 and 2,
respectively. It is seen that the quantum results are well fitted
by the semiclassical Fano factors. According to both analyses, the
third-harmonic mode has the most suppressed photocount noise. }
\label{bajefg08} \vspace*{-5mm}
\end{figure}

For better comparison of theoretical predictions for different
order processes, we have plotted the quantum Fano factors for
both interacting modes in the no-energy-transfer regime with
$N=2-5$ and $r=5$ in Fig. \ref{bajefg07}. One can see that all
curves start from $F^{\rm Q}_{1,N}\left( 0\right) =1$ for the
input coherent fields and become quasi-stationary after some
relaxations. The quantum and semiclassical Fano factors coincide
for high-intensity fields and longer times, specifically for
$t\geq 50/(\overline{\Omega }g)$, where $\overline{\Omega }$ will
be defined later by Eq. (\ref{N54}).  In Fig. \ref{bajefg07}, we
observe that all fundamental modes remain super-Poissonian
($F^{\rm Q}_{1}\left( t\right) >1$), whereas the $N$th harmonics
become sub-Poissonian ($F^{\rm Q}_{N}\left( t\right) <1$). The
most suppressed noise is observed for the third harmonic with the
Fano factor $F^{\rm Q}_{3}\approx 0.81$. In Fig. \ref{bajefg07},
we have included the predictions of the classical trajectory
method (plotted by dotted lines) to show that they properly fit
the exact quantum results (full curves) for the evolution times
$t\geq 50/(\overline {\Omega }g)$. The small residual differences
result from the fact that the amplitude $r$ was chosen to be
relatively small ($r=5$). This value does not precisely fulfill
the condition $r\gg 1$. We have taken $r=5$ as a compromise
between the asymptotic value $r\rightarrow \infty $ and
computational complexity to manipulate the matrices of dimensions
$1000\times 1000$. Unfortunately, we cannot increase amplitude
$r$ arbitrary due to computational limitations.
\begin{table}[hb]
\caption{Quasi-stationary values of the quantum Fano factors
$F^{\rm Q}_{1}$ and their semiclassical approximations $F^{\rm
S}_{1}$, given by (\ref{N54}), for the fundamental mode in
$N$th-harmonic generation with $N=1-5$ in the no-energy-transfer
regime. The values of $F^{\rm Q}_{1}$ are calculated for
$r=r_{N}=5$. }
\begin{tabular*}{\textwidth}{@{\extracolsep{\fill}}cccc}
\hline $N$ & $F^{\rm Q}_{1}$ & $F^{\rm S}_{1}$ & $(F^{\rm
Q}_{1}-F^{\rm S}_{1})/F^{\rm Q}_{1}$   \cr \hline
1 & 1 & 1 & 0   \\
2 & 1.5029291 & 3/2 & 0.0020   \\
3 & 1.8202032 & 29/16 & 0.0042   \\
4 & 2.0323293 & 101/50 & 0.0061   \\
5 & 2.1830414 & 13/6 & 0.0075 \\
\hline
\end{tabular*}
\label{bajetb01}
\end{table}
\begin{table}[hb]
\caption{Same as in table 1, but for the $N$th harmonic mode;
$F^{\rm S}_{N}$ are calculated from (\ref{N55}). }
\begin{tabular*}{\textwidth}{@{\extracolsep{\fill}}cccc}
\hline
$N$ & $F^{\rm Q}_{N}$ & $F^{\rm S}_{N}$ & $|F^{\rm Q}_{N}
-F^{\rm S}_{N}|/F^{\rm Q}_{N}$   \\
\hline
1 & 1 & 1 & 0   \\
2 & 0.83228800 & 5/6 & 0.0013  \\
3 & 0.81125970 & 13/16 & 0.0015  \\
4 & 0.81924902 & 41/50 & 0.00092  \\
5 & 0.83331127 & 5/6 & 0.000026 \\
\hline
\end{tabular*}
\label{bajetb02}
\end{table}
Numerical values of the quantum Fano factors in comparison with
their semiclassical approximations for the fundamental mode,
given by Eq. (\ref{N56}), are presented in their dependence on $N$
in table \ref{bajetb01} and Fig. \ref{bajefg08}(a).  Analogously,
those values for harmonics are presented in Fig.
\ref{bajefg08}(b) and table \ref{bajetb02} as calculated by the
numerical quantum method and from analytical semiclassical formula
(\ref{N57}). It is seen that the approximate predictions of the
Fano factors, according to (\ref{N56}) and (\ref{N57}), fit very
well the values obtained by applying the numerical quantum
method.  In fact, the differences between the approximate and
exact values are hardly visible on the scale of Fig.
\ref{bajefg08}.  Nevertheless, some small ($<1\%$) differences in
$F^{\rm Q}_{1,N}$ (see tables \ref{bajetb01} and \ref{bajetb02})
can be explained by the fact that the value of $r$ for numerical
analysis was chosen too small.

\subsection{Classical analysis}
\label{sect32}

Our classical analysis of higher-harmonic generation follows the
same method as described in Sect. \ref{sect23}. The classical
model of the $N$th-harmonic generation can be described by
\begin{equation}
{\cal H}= g\left[
\alpha_{1}^{N}\alpha_{N}^{*}+(\alpha_{1}^*)^{N}\alpha_{N}\right] ,
\label{N35}
\end{equation}
which in a special case of $N=2$ goes over into Eq. (\ref{N11}).
In Eq. (\ref{N35}), $\alpha_{1}$ ($\alpha_{N}$) is the complex
amplitude of the fundamental ($N$th-harmonic) mode. Hamiltonian
(\ref{N35}) leads to the pair of complex differential equations
\cite{Boy91}
\begin{eqnarray}
\dot{\alpha} _{1}&=&-{\rm i}gN\alpha _{1}^{\ast N-1}\alpha _{N},
\nonumber\\
\dot{\alpha} _{N}&=&-{\rm i}g\alpha _{1}^{N}, \label{N36}
\end{eqnarray}
On introducing real amplitudes and phases, $\alpha _{k}=r_{k}{\rm
e}^{{\rm i}\phi _{k}}$, (\ref{N36}) can be transformed into the
system of three real equations:
\begin{eqnarray}
\dot{r}_{1} &=&-gNr_{1}^{N-1}r_{N}\sin \theta ,
\nonumber\\
\dot{r}_{N}&=&gr_{1}^{N}\sin \theta ,
\nonumber\\
\dot{\theta} &=&g\left( r_{1}^{N}/r_{N}-N^{2}r_{1}^{N-2}r_{N}\right) \cos
\theta ,
\label{N37}
\end{eqnarray}
where $\theta =N\phi _{1}-\phi _{N}$ is the phase mismatch. Equations
(\ref{N37}) have two integrals of motion:
\begin{eqnarray}
E&=&r_{1}^{2}+Nr_{N}^{2}=n_{1}+Nn_{N},
\nonumber\\
\Gamma&=&r_{1}^{N}r_{N}\cos \theta. \label{N38}
\end{eqnarray}
On extraction of $r_{1}$ and $\theta $ from Eq. (\ref{N37}), we
find equation for the amplitude $r_{N}$:
\begin{eqnarray}
\left( r_{N}\dot{r}_{N}/g\right) ^{2}+\Gamma ^{2}=r_{N}^{2}\left(
E-Nr_{N}^{2}\right) ^{N}
\label{N39}
\end{eqnarray}
or its simpler form for the intensity $n_{N}=r_{N}^{2}$:
\begin{equation}
\left( \dot{n}_{N}/2g\right) ^{2}=n_{N}\left( E-Nn_{N}\right) ^{N}-\Gamma
^{2}.  \label{N40}
\end{equation}
The general solution for $n_{N}\left( t\right) $ is a periodic
function oscillating between the values $n_{\min }$ and $n_{\max
}$. The solution can be given in terms of the Jacobi elliptic
functions \inxx{Jacobi elliptic function} for $N=2$ and $N=3$,
and in terms of hyperelliptic functions for $N>3$.

One elementary solution of set of Eqs. (\ref{N37}) is obtained for
the zero initial phase mismatch $\theta =0$ and the initial
amplitudes satisfying the condition $r_{1}=Nr_{N}$. The solution
reads as
\begin{eqnarray}
\alpha _{1}\left( t\right) &=&r_{1}\exp (-{\rm i}gtr_{1}^{N-1}),
\nonumber\\ \alpha _{N}\left( t\right) &=&r_{N}\exp \left( -{\rm
i}Ngtr_{1}^{N-1}\right), \label{N41}
\end{eqnarray}
which corresponds to the \inx{no-energy-transfer regime}, since
the amplitude and energy in both the interacting modes remain
constant~ $n_{1}\left( t\right) =\left| \alpha _{1}\left(
t\right) \right| ^{2}=r_{1}^{2}$ and $n_{N}\left( t\right)
=\left| \alpha _{N}\left( t\right) \right| ^{2}=r_{N}^{2}$
\cite{Jex97}.

\subsection{Classical trajectory analysis}
\label{sect33} \inxx{classical trajectory analysis}

The results of Sect. \ref{sect32} can be used in the method of
classical trajectories in analogy with the technique described in
Sect. \ref{sect23}. We need to express the trajectories in their
dependence on small noise parameters $x_{k}$ and $y_{k}.$ The
integrals of motion, given by (\ref{N38}), can be expressed in a
form of corrections in successive powers of large $r$:
\begin{eqnarray}
E=N\left( N+1\right) r^{2}+\Delta E_{1}+\Delta E_{0},
\label{N42}
\end{eqnarray}
where
\begin{eqnarray}
\Delta E_{1} &=&2N\left( x_{1}+x_{N}\right) r,
\nonumber\\
\Delta E_{0} &=&x_{1}^{2}+y_{1}^{2}+N\left( x_{N}^{2}+y_{N}^{2}\right) ,
\label{N43}
\end{eqnarray}
and
\begin{equation}
\Gamma =N^{N}r^{N+1}+\Delta \Gamma _{N}+\Delta \Gamma _{N-1}+\Delta \Gamma
_{N-2}+\cdots ,
\label{N44}
\end{equation}
where
\begin{eqnarray}
\Delta \Gamma _{N} &=&\left( x_{1}+x_{N}\right) \left( Nr\right) ^{N},
\nonumber\\
\Delta \Gamma _{N-1} &=&\left[ \frac{N-1}{2}\left(
x_{1}^{2}-y_{1}^{2}\right) +N\left( x_{1}x_{N}+y_{1}y_{N}\right) \right]
\left( Nr\right) ^{N-1}.
\label{N45}
\end{eqnarray}
The lower-order terms $\Delta \Gamma _{N-2},$ $\Delta \Gamma
_{N-3},...$ can be neglected in further considerations. On
assumption of high-intensity fields ($r\gg 1$), we can substitute
\begin{equation}
n_{N}=\frac{E}{N\left( N+1\right) }+\epsilon ,
\label{N46}
\end{equation}
where $\epsilon $ is a small correction of stationary value.
Then, r.h.s. of (\ref{N40}) can be rewritten as
\begin{equation}
n_{N}\left( E-Nn_{N}\right) ^{N}-\Gamma ^{2}\approx \frac{N^{N}}{2\left(
N+1\right) ^{N-2}}E^{N-1}\left( A^{2}-\epsilon ^{2}\right)
\label{N47}
\end{equation}
on omission of higher-order terms involving $\epsilon ^{3}$, $\epsilon
^{4},\cdots $. One arrives at simple equation
\begin{eqnarray}
\left( \frac{\dot{\epsilon}}{2g}\right) ^{2}=\frac{N^{N}}{2\left( N+1\right)
^{N-2}}E^{N-1}\left( A^{2}-\epsilon ^{2}\right) .
\label{N48}
\end{eqnarray}
Thus, the solution of (\ref{N40}) reads as
\begin{equation}
n_{N}\left( t\right) =r^{2}+B+A\sin \Omega gt,
\label{N49}
\end{equation}
where the frequency $\Omega $ is given by
\begin{eqnarray}
\Omega =\sqrt{\frac{2N^{N}E^{N-1}}{\left( N+1\right) ^{N-2}}}
\label{N50}
\end{eqnarray}
and
\begin{eqnarray}
&&A=\frac{r}{N+1}\sqrt{4\left( x_{1}-Nx_{N}\right) ^{2}+2N\left( N+1\right)
\left( y_{1}-y_{N}\right) ^{2}}
\nonumber\\
&&B=\frac{\Delta E_{1}}{N\left( N+1\right) }=\frac{2}{N+1}r\left(
x_{1}+x_{N}\right) .
\label{N51}
\end{eqnarray}
From (\ref{N38}), a result similar to (\ref{N49}) is obtained for the
fundamental mode:
\begin{equation}
n_{1}\left( t\right) =E-Nn_{N}\left( t\right) =N^{2}r^{2}+N^{2}B-NA\sin
\Omega gt.  \label{N52}
\end{equation}
It is seen that both solutions (\ref{N49}) and (\ref{N52}) are
given in a form of large constants weakly perturbed by harmonic
function.

Now, on applying the classical trajectory method, one should perform
averaging over all solutions (\ref{N49}) and (\ref{N52}) to calculate the
required statistical moments. Here, we calculate the first and second-order
field-intensity moments necessary for determination of the Fano factors. The
mean intensities of the fundamental and harmonic modes are simply given by
$\overline{n}_{1}=N^{2}r^{2}$ and $\overline{n}_{N}=r^{2}$, respectively. The
second-order moments of field intensity are found to be
\begin{eqnarray}
\overline{n_{1}^{2}}&=&N^{4}r^{4}+N^{4}\overline{B^{2}}+\frac{1}{2}N^{2}
\overline{A^{2}}, \nonumber\\
\overline{n_{N}^{2}}&=&r^{4}+\overline{B^{2}}+\frac{1}{2}
\overline{A^{2}}. \label{N53}
\end{eqnarray}
in terms of $\overline{A^{2}}=r^{2}(2N^{2}+N+1)/(N+1)^{2}$ and
$\overline {B^{2}}=2r^{2}/(N+1)^{2}$. We note that $\overline{B}$
vanishes. The term $\overline{\sin ^{2}\Omega gt}$ can simply be
estimated with $1/2$ for sufficiently long time $t$, when
$n_{k}\left( t\right) $ and $F^{\rm Q}_{k}\left( t\right) $
become quasi-stationary. The relaxation in $n_{k}\left( t\right) $
and $F^{\rm Q}_{k}\left( t\right) $ is observed for short times
$t$ due to the presence of harmonic sine function and residual
phase synchronization. The mean value of the frequency
(\ref{N50}), given by
\begin{equation}
\overline{\Omega}\approx \sqrt{2N\left( N+1\right) }
\left( Nr\right)^{N-1},
\label{N54}
\end{equation}
enables estimation of the oscillation period $T_{\rm osc}=2\pi
/\overline{\Omega }$, whereas the standard deviation
\begin{equation}
\Delta \Omega \approx \sqrt{2N\left( N+1\right) }N^{N-1}r^{N-2}\frac{N-1}
{N+1}
\label{N55}
\end{equation}
determines the duration $T_{\rm rel}=2\pi /\Delta \Omega $ of
relaxation. By comparing the characteristic times $T_{\rm osc}$
and $T_{\rm rel}$, one finds that the evolution time can be
scaled by $\tau =\overline{\Omega }gt$ to synchronize optimally
the oscillations of the exact quantum solutions for different $N$.
These synchronized oscillations of the Fano factors are clearly
presented in Fig. \ref{bajefg07}.

Finally, we arrive at the semiclassical Fano factors \inxx{Fano
factor; semiclassical}
\begin{eqnarray}
F^{\rm S}_{1} &=&\frac{1}{2}\frac{6N^{2}+N+1}{\left( N+1\right)
^{2}},
\label{N56}\\
F^{\rm S}_{N} &=&\frac{1}{2}\frac{2N^{2}+N+5}{\left( N+1\right)
^{2}}, \label{N57}
\end{eqnarray}
which are the compact-form analogs of the quantum Fano factors.
The semiclassical Fano factors for the fundamental and higher
harmonics for various values of $N$ are listed in tables
\ref{bajetb01} and \ref{bajetb02}, and plotted in Figs.
\ref{bajefg07}(a) and \ref{bajefg07}(b), respectively.

Our solutions (\ref{N56}) and (\ref{N57}) reduce to the results
derived in Ref. \cite{Baj99} for $N=2$, and those of Ref.
\cite{Baj00} for $N=3$. By analyzing (\ref{N57}), we find that
higher harmonics evolve into quasi-stationary sub-Poissonian
states ($F^{\rm S}_{N}<1$) for any $N>1$. Except for second
harmonic, the photocount noise reduction in higher harmonics
becomes less effective with increasing $N $. Thus, the deepest
noise reduction occurs for the third harmonic as described by the
Fano factor $F^{\rm S}_{3}=\frac{13}{16}=\allowbreak 0.8125$. The
photocount noise reductions for the second and fifth harmonics
are predicted to be the same, although the quantum analysis (see
table \ref{bajetb02}) reveals that they differ slightly ($<1\%$).
As comes from (\ref{N56}), the fundamental mode has solely the
super-Poissonian photocount statistics ($F^{\rm S}_{1}>1$) with
noise monotonically growing in $N$ for the no-energy-transfer
regime. For $N=1$, the process is linear and no change in the
photon statistics occurs. The interacting modes remain coherent
with the unit Fano factors for both modes. It is worth noting
that qualitatively different photocount statistics of the
fundamental mode is observed in the short-interaction regime as
given by Eqs. (\ref{N06}) and (\ref{N33})-(\ref{N34}).

We have shown, in agreement with the results presented in \cite{Baj99},
that the method of classical trajectories gives very good predictions in
the case of strong-field interactions (i.e., for the photon numbers larger
than 10).  The calculation speed of the method does not depend on numbers
of interacting photons. But better approximation is achieved with the
increasing number of photons. Thus, the method is very fast and
significantly simplifies the tedious exact quantum calculations.

\section{Conclusion}

We have presented quantum, classical and semiclassical
descriptions of second and higher harmonic generations. We have
demonstrated that these processes can be a source of
sub-Poissonian light. On testing different coherent input
amplitudes and phases in order to minimize the Fano factor, we
have discovered a quantum regime for which the long-interaction
output is generated with the quasi-stationary sub-Poissonian
photocount noise \cite{Baj99,Baj00,Baj00a}. The regime occurs if
the initial coherent state amplitudes are related by $|\alpha
_{1}| =N|\alpha _{N}|$ and  ${\rm Arg}(\alpha_{N})\simeq N{\rm
Arg}(\alpha_{1})$. This is a quantum analog of the
no-energy-transfer regime \cite{Jex97} known in classical
nonlinear optics as an evolution exhibiting no-energy transfer
between the interacting modes. The intensities of both modes
remain quasi-constant in time during the interaction. Obviously,
in a quantum analysis some small energy fluctuations between
modes are observed as a consequence of vacuum fluctuations.
However, the influence of energy fluctuations can be neglected
for strong fields.

We have proved that in the no-energy-transfer regime, the
fundamental mode evolves into a quasi-stationary state with the
super-Poissonian ($F^{\rm Q}_{1}>1$) photocount statistics,
whereas the $N$-th harmonic goes over into a sub-Poissonian
($F^{\rm Q}_{N}<1)$ quasi-stationary state. We have found that
the most suppressed photocount noise is obtained for the third
harmonic as described by the quantum Fano factor $F^{\rm
Q}_{3}=0.811\cdots$. Good analytical predictions of the quantum
Fano factors for both the fundamental and harmonic modes ($F^{\rm
S}_{3}=13/16=0.8125$) were obtained under the semiclassical
approximation in the strong-field limit.

\begin{acknowledgments}
J.B. thanks Prof. Jan Pe\v{r}ina and Dr. Ond\v{r}ej Haderka for
interesting discussions. A.M. wishes to thank Professors Nobuyuki
Imoto and Masato Koashi for their hospitality and stimulating
research at SOKEN. J.B. was supported by the Ministry of
Education of Czech Republic under Projects  VS96028, CEZ J14/98
and LN00A015, and by the Grant Agency of Czech Republic (No.
202/00/0142). A.M. was supported by the Japan Science and
Technology Corporation (JST-CREST).
\end{acknowledgments}


\end{document}